\documentclass[preprint,showpacs,aps,preprintnumbers,amsmath,amssymb]{revtex4-1}

\usepackage{graphicx,epsfig}
\usepackage{dcolumn}
\usepackage{bm}
\usepackage{xcolor}
\newcommand{\be}{\begin{equation}}
\newcommand{\ee}{\end{equation}}
\newcommand{\bse}{\begin{subequations}}
\newcommand{\ese}{\end{subequations}}
\newcommand{\bary}{\begin{eqnarray}}
\newcommand{\eary}{\end{eqnarray}}
\newcommand{\bwt}{\begin{widetext}}
\newcommand{\ewt}{\end{widetext}}

\begin{document}


\title{The VHE SED modelling of Markarian 501 in 2009}
\author{Sarira Sahu$^a$ }
\email{sarira@nucleares.unam.mx}
\author{Carlos E. L\'opez Fort\'in$^a$} 
\email{carlos.fortin@correo.nucleares.unam.mx}
\author{Miguel E. Iglesias Mart\'inez$^{b,c}$}
\email{migueliglesias2010@gmail.com}
\author{Shigehiro Nagataki$^{d,e}$}
\email{shigehiro.nagataki@riken.jp}
\author{Pedro Fern\'andez de C\'ordoba$^c$}
\email{pfernandez@mat.upv.es}

\affiliation{
$^a$Instituto de Ciencias Nucleares, Universidad Nacional Aut\'onoma de M\'exico, 
Circuito Exterior, C.U., A. Postal 70-543, 04510 Mexico DF, Mexico\\
$^b$Departamento de Telecomunicaciones, Universidad de Pinar del R\'io,
Pinar del R\'io, Cuba, Martí  \# 270, CP: 20100\\
$^c$Instituto Universitario de Matemática Pura y Aplicada, Universitat Polit\`ecnica de València (UPV), Camino de Vera s/n, 46022 Valencia, España\\
$^d$Astrophysical Big Bang Laboratory, RIKEN, Hirosawa, Wako, Saitama 351-0198, Japan\\
$^{e}$Interdisciplinary Theoretical \& Mathematical Science (iTHEMS),
RIKEN, Hirosawa, Wako, Saitama 351-0198, Japan
}

\begin{abstract}

The high energy blazar, Markarian 501 was observed 
as a part of multi-instrument and multiwavelength campaign spanning
the whole electromagnetic spectrum for 4.5 months during March 15 to
August 1, 2009. On May 1, Whipple 10m telescope observed a very strong
$\gamma$-ray flare in a time interval of about 0.5 h. Apart from this flare, high
state and low state emissions were also observed by Whipple, VERITAS
and MAGIC telescopes. Using the photohadronic model and accounting for the
absorption of the extragalactic background light (EBL) to these
very high energy $\gamma$-rays, excellent fits are obtained for the 
observed spectra. We have shown that the intrinsic spectrum for low state
emission is flat, however, for high and very high states this is a
power-law with slowly increasing exponent.

\end{abstract}
\maketitle

\section{Introduction}

Markarian 501 (Mrk 501) is a high energy peaked blazar (HBL) at a redshift of
$z=0.034$ and one of the brightest extragalactic sources in the
X-ray/TeV sky. Mrk 501 is identified as a very high energy
(VHE, $> 100$ GeV) emitter by the Whipple telescope in 1996\cite{Abdo:2009wu}. Since its
discovery, it has been the subject of extensive studies in
multiwavelengths as it has undergone many major outbursts on long time
scales and rapid flares on short time scales mostly in the X-rays and
TeV energies\cite{Pian:1998hh,Krawczynski:2003fq,Tavecchio:2001,Ghisellini:1998it,Sambruna:2000ic,Gliozzi:2006qq,Villata:1999,Katarzynski:2001,Aharonian:1998jw,Aharonian:1999vy,Aharonian:2000xr}. 
As a part of multiwavelength campaign covering a period
of 4.5 months from March 15 to August 1, 2009, Mrk 501 was observed by
multi-instruments ($\sim 30$ different instruments) covering the
entire electromagnetic spectrum\cite{Aliu:2016kzx,Ahnen:2016hsf}. In the optical and radio bands the
flux was found to be almost constant and in the UV band it had some
variation, however, during the VHE flare the flux was constant. Around
the epoch of VHE flaring only, the X-ray light curve exhibited variation.
In the $\gamma$-ray range from 0.1 GeV to 20 TeV, it
was observed by {\it Fermi-LAT}, MAGIC, VERITAS and Whipple 10m
telescopes. The $\gamma$-ray telescopes observed two episodes of flaring, one on 1st
May (MJD 54952) and another on 22nd May (MJD 54973). 
In the VHE domain, statistically significant variation in
the flux was observed by all the instruments.


The VHE flare of May 1, 2009 was first observed by the Whipple 10m telescope
when a high emission state was detected above 300 GeV\cite{Aliu:2016kzx,Ahnen:2016hsf}. A sudden
increase in the flux in the first 0.5 h (MJD 54952.31 to MJD 54952.37)
by about one order of magnitude 
compared to the average flux measured during the full campaign was
recorded\cite{Aliu:2016kzx}. Following an alert by the Whipple telescope, 1.4 h 
later (MJD 54952.41), VERITAS continued simultaneous observation with
the Whipple and detected elevated level of flux without statistically
significant variation in it during the full period of observation (MJD
54952.41 to MJD 54952.48). Both Whipple and VERITAS observed elevated
flux level in the source until MJD 54955. 

The MAGIC collaboration participated in this campaign with a
single telescope only. Also due to upgrading of the telescope, data
were not taken during the whole campaign period. However, on May 22
(MJD 54973), the MAGIC telescope observed the blazar for 1.7 h and
an elevated VHE flux (a VHE flare) $\sim 3$ times the low flux level 
has been recorded\cite{Ahnen:2016hsf}. During this flaring period, no significant
intra-night variability was observed. The VHE $\gamma$-ray spectra
observed during the 4.5 months period by  
different instruments are summarized in Table \ref{table1} and also
the EBL corrected spectra are given in ref.\cite{Ahnen:2016hsf}. 

In a recent paper Sahu et al, have explained the VHE flare of May 1 well using the
photohadronic scenario by taking into account the EBL correction to
the spectrum\cite{Sahu:2016mww,Sahu:2019lwj}. However, the SSC spectrum used to explain the flare data was
non-simultaneous with the VHE spectrum. Apart from this, 
the comparison between the 
high energy spectrum of refs.\cite{Aliu:2016kzx} and\cite{Ahnen:2016hsf} shows that
there is a difference in their spectral behavior.


In the present work, we use the same photohadronic
model to explain the spectra from low to very high states of Mrk 501 during the 4.5
months campaign.
For the first time we have shown here that, very high, high and low
states VHE flaring from Mrk 501 can be explained very well
and simultaneously using the photohadronic scenario.

\begin{table}
\centering
\caption{Summary of the $\gamma$-ray spectra measured by different instruments in
  different time intervals. The time is shown in Modified Julian Date (MJD)\cite{Ahnen:2016hsf}.
} 
\label{table1}
\begin{tabular*}{\columnwidth}{@{\extracolsep{\fill}}llll@{}}
\hline
\multicolumn{1}{@{}l}{Instrument} &Flux state & Period (in MJD unit) \\
\hline
{\it Whipple} & low & 54936 -- 54951\\
VERITAS & low & 54907 -- 55004\\
MAGIC & low & 54913 -- 55038\\
\hline
{\it Whipple} & high & 54952.41--54955\\
VERITAS & high & 54952.41--54955\\ 
MAGIC & high & 54973\\
\hline
{\it Whipple} & very high &54952.35-- 54952.41\\
\hline
\end{tabular*}
\end{table}

\section{The Model}

Blazars are a sub class of AGN and have non thermal spectra. Rapid
variability is observed in their entire electromagnetic spectra. This
implies that the observed photons originate within the highly relativistic
jets oriented very close to the observers line of sight\cite{Urry:1995mg,Acciari:2010aa}.
In the $\nu-\nu F_{\nu}$ plane, their spectral energy distributions
(SEDs) have a double peak structure. These two peaks are explained by
leptonic models. In this scenario, relativistic electrons
radiate synchrotron photons in the jet magnetic field giving the first
peak. The Compton scattering of the high energy electrons with the
self-produced synchrotron photons (synchrotron self Compton, i.e. SSC)
gives rise to the second peak\cite{Dermer:1993cz,Sikora:1994zb}.
In this model, the emitting region is a blob
with comoving radius $R'_b$ (the $^{\prime}$ implies the jet
is in comoving frame and without prime are in the observer frame), moving with a bulk Lorentz
factor $\Gamma$ and a Doppler factor ${\cal
  D}$\cite{Ghisellini:1998it,Krawczynski:2003fq} and for HBLs
$\Gamma\simeq {\cal D}$. 

The photohadronic model of Sahu et al.\cite{Sahu:2015tua,Sahu:2016bdu,
Sahu:2016mww,Sahu:2019lwj} can explain the flaring events from HBLs very well and
it  relies on
the standard interpretation of the leptonic model discussed above to explain both low and
high energy peaks by synchrotron and SSC photons respectively as in the
case of any other AGN. 

A double jet structure scenario is assumed to explain the multi-TeV
emission from the HBLs. The jet which is compact and smaller in size
($R'_f$) is enclosed within the bigger cone of size
$R'_b$ ($R'_f < R'_b$) and both have almost the same bulk Lorentz factor $\Gamma$.
As the compact jet is hidden we cannot probe directly its density
$n'_{\gamma, f}$. However, the photon density in the outer region $n'_{\gamma}$ is
known from the observed flux. The photon density in the inner jet will
decrease due to its adiabatic expansion when it crosses into the outer
region. To connect the inner and the outer jet regions, a scaling
behavior of their photon densities is proposed as
\be
\frac{n'_{\gamma, f}(\epsilon_{\gamma_1})}
{n'_{\gamma, f}(\epsilon_{\gamma_2})} \simeq \frac{n'_\gamma(\epsilon_{\gamma_1})}
{n'_\gamma(\epsilon_{\gamma_2})}.
\label{densityratio}
\ee 
In the above equation the right hand side is known, however, the left
hand is unknown. So we can use this relation to express the unknown
photon density in the inner region in terms of the photon density in
the outer region.
The Fermi accelerated protons in the inner jet region have a power-law
spectrum given by
\be
dN/dE_p \propto E^{-\alpha}_p, \, \, \, \, \, \alpha \ge 2. 
\ee
These protons interact with the photons in the inner jet region to
produce the $\Delta$-resonance. The subsequent decay of the
$\Delta$-resonance to $\gamma$-rays and neutrinos takes place by
intermediate $\pi^0$ and $\pi^+$ states respectively.

The resonance process $p\gamma\rightarrow \Delta$ gives the
kinematical condition
\be
E_{\gamma}  \epsilon_{\gamma} \simeq 0.032 \frac{{\cal
    D}^2}{(1+z)^2}\, GeV^2,
\label{KinCon}
\ee
where $E_{\gamma}$ is the observed VHE $\gamma$-ray energy, 
$\epsilon_{\gamma}$ is the the background seed photon energy and $z$
 is the redshift of the HBL. In the above process, the VHE photon
 carries about 10\% of the proton energy ($E_{p} \simeq 10E_{\gamma}$).

The observed VHE $\gamma$-ray flux is proportional to $n'_{\gamma,f}$
and proton flux $E^2_p\,dN/dE_p$.
In a traditional jet scenario, the photohadronic process is inefficient
due to the low photon density and
super-Eddington luminosity\cite{Cao:2014nia,Zdziarski:2015rsa} in
proton is needed to explain the multi-TeV emission.
Using the scaling condition of Eq. (\ref{densityratio}), the photon
density in the inner region can be expressed in terms of the observed
flux\cite{Sahu:2016mww,Sahu:2019lwj}. 

The observed range of multi-TeV $E_{\gamma}$ corresponds to the range
of $\epsilon_{\gamma}$ which lies in the low energy tail region of the
SSC spectrum. Using the relation
\be
n'_{\gamma}(\epsilon_{\gamma})=\eta \left ( \frac{d_L}{R'_b} \right
)^2 \frac{1}{(1+z)} \frac{\Phi_{SSC}(\epsilon_{\gamma})}{{\cal
    D}^{2+\kappa}\, \epsilon_{\gamma}},
\label{photondensity}
\ee
where  $\eta$ is the efficiency of the SSC process and we take
$\eta=1$ for 100\% efficiency. The parameter $\kappa=0(1)$ 
corresponds to continuous (discrete) blazar jet and $d_L\simeq 156$ Mpc is the
luminosity distance of Mrk 501.
So $F_{\gamma}\propto n'_{\gamma}$
implies $F_{\gamma}\propto \Phi_{SSC}$.
Also remember that, high energy gamma-rays get
attenuated on their way to Earth. So the observed multi-TeV spectrum
has to be corrected for the extragalactic background light (EBL)
absorption. Taking the EBL correction into account, the observed multi-TeV flux can be given as
\be
F(E_{\gamma})=A_{\gamma} \Phi_{SSC}(\epsilon_{\gamma} )\left (
  \frac{E_{\gamma}}{TeV}  \right)^{-\alpha+3}
e^{-\tau_{\gamma\gamma}(E_{\gamma},z)},
\label{modifiedsed}
\ee
where $A_{\gamma}$ is the normalization constant which can be fixed
from the observed VHE data and $\tau_{\gamma\gamma}$ is the
energy and redshift dependent optical depth for the pair creation
process. The low energy tail region of the SSC spectrum can
be expressed as $\Phi_{SSC} \propto \epsilon^{\beta}_{\gamma}$ with $\beta > 0$,
irrespective of whether it is in quiescent state or in flaring
state. However, the value of $\beta$ does differ in different
states. Using Eq. (\ref{KinCon}), we can express
\be
\Phi_{SSC}(\epsilon_{\gamma}) =\Phi_0\, \left (\frac{E_{\gamma}}{
      TeV}\right )^{-\beta}.
\label{phi_power_law}
\ee
Substituting Eq. (\ref{KinCon}) in Eq. (\ref{modifiedsed}), the observed flux can be expressed as 
\be
F_{\gamma} (E_{\gamma}) = F_{\gamma,
  in}(E_{\gamma})\,e^{-\tau_{\gamma\gamma}(E_{\gamma},z)}, 
\label{obsflux}
\ee
where the intrinsic flux is defined as
\be
F_{\gamma, in}(E_{\gamma}) =F_0 \left (\frac{E_{\gamma}}{TeV}
\right )^{-\alpha-\beta+3}.
\ee
The value of $\beta$ is obtained from the low energy tail region of the SSC SED
of a given leptonic model. So the only free parameter here is the
proton spectral index $\alpha$.

During the VHE $\gamma$-ray emission state the flux from the jet can
be as high as $F_{Edd}/2$, where $F_{Edd}$ is the Eddington flux. Also
the highest energy protons emitted in the emission process should satisfy
$F_p < F_{Edd}/2$ which will constrain the optical depth
$\tau_{p\gamma}$ for the $p\gamma\rightarrow\Delta$ process and
consequently the $n'_{\gamma, f}$ in the inner region. 
The hidden jet lies between $R_s$
(Schwarzschild radius) and $R'_b$ so that photon density in the inner region is high \cite{Sahu:2016mww,Sahu:2019lwj}.

\begin{figure}
{\centering
\resizebox*{0.8\textwidth}{0.5\textheight}
{\includegraphics{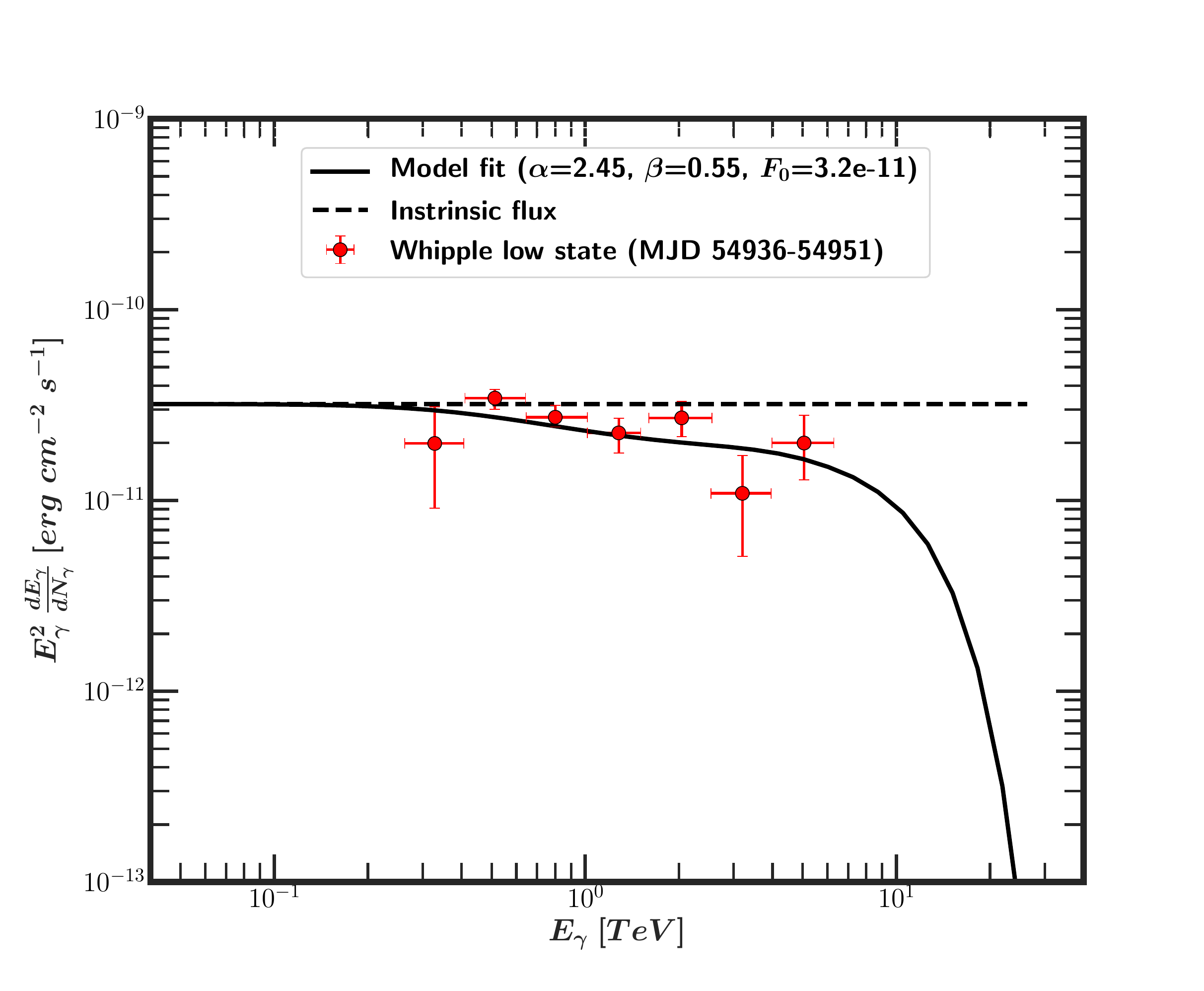}}
\par}
\caption{
Whipple low state spectrum (MJD 54936 to 54951) is fitted with the
photohadronic model.
In all the figures the normalization constant $F_0$ is given in units
of $erg\, cm^{-2}\, s^{-1}$, and the dashed curve corresponds to
intrinsic flux.
\label{fig:figure1}
}
\end{figure}
\begin{figure}
{\centering
\resizebox*{0.8\textwidth}{0.5\textheight}
{\includegraphics{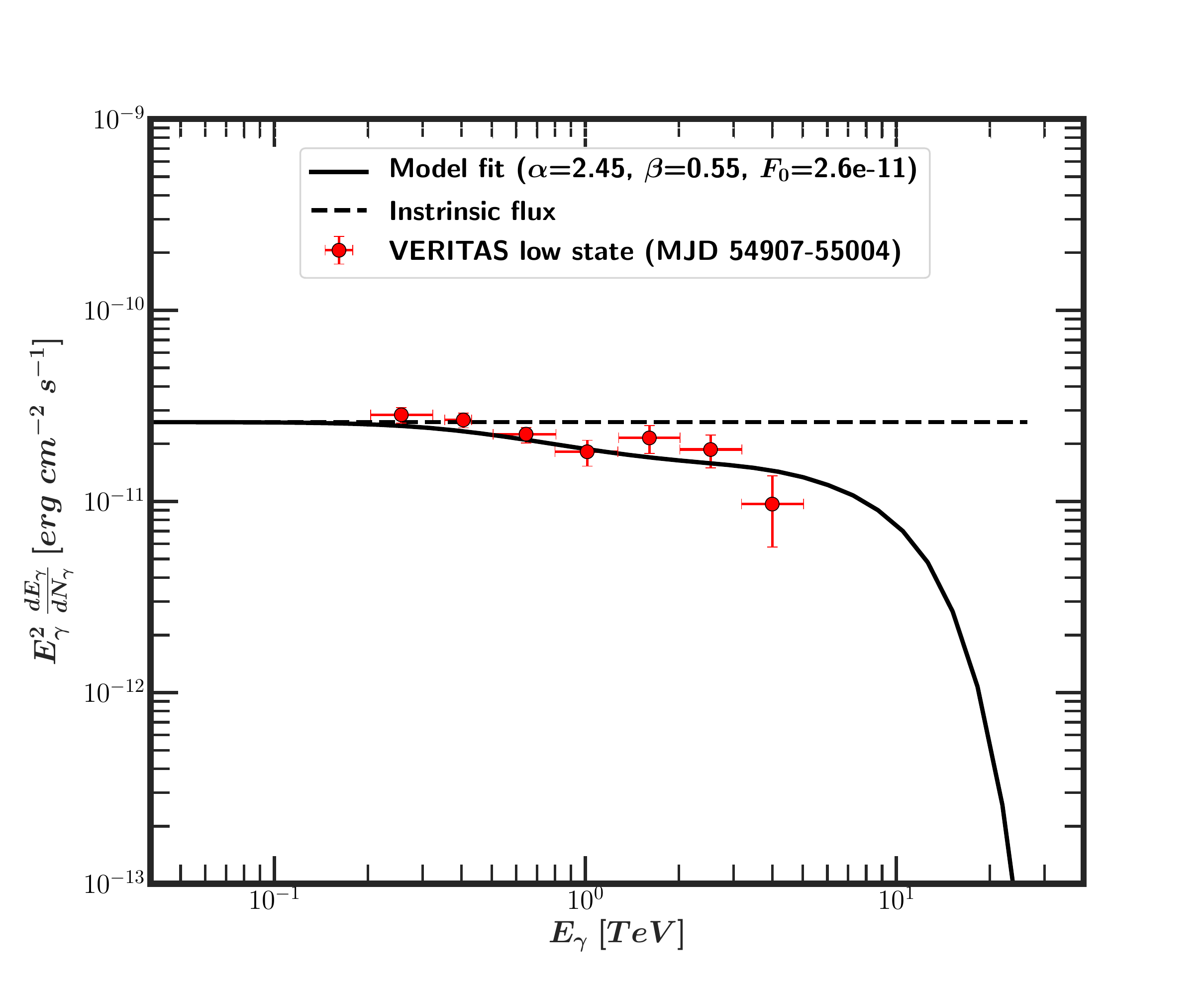}}
\par}
\caption{
VERITAS low state spectrum (MJD 54907 to MJD 55004) is fitted with the
photohadronic model. The corresponding intrinsic flux is also given.
\label{fig:figure2}
}
\end{figure}
\begin{figure}
{\centering
\resizebox*{0.8\textwidth}{0.5\textheight}
{\includegraphics{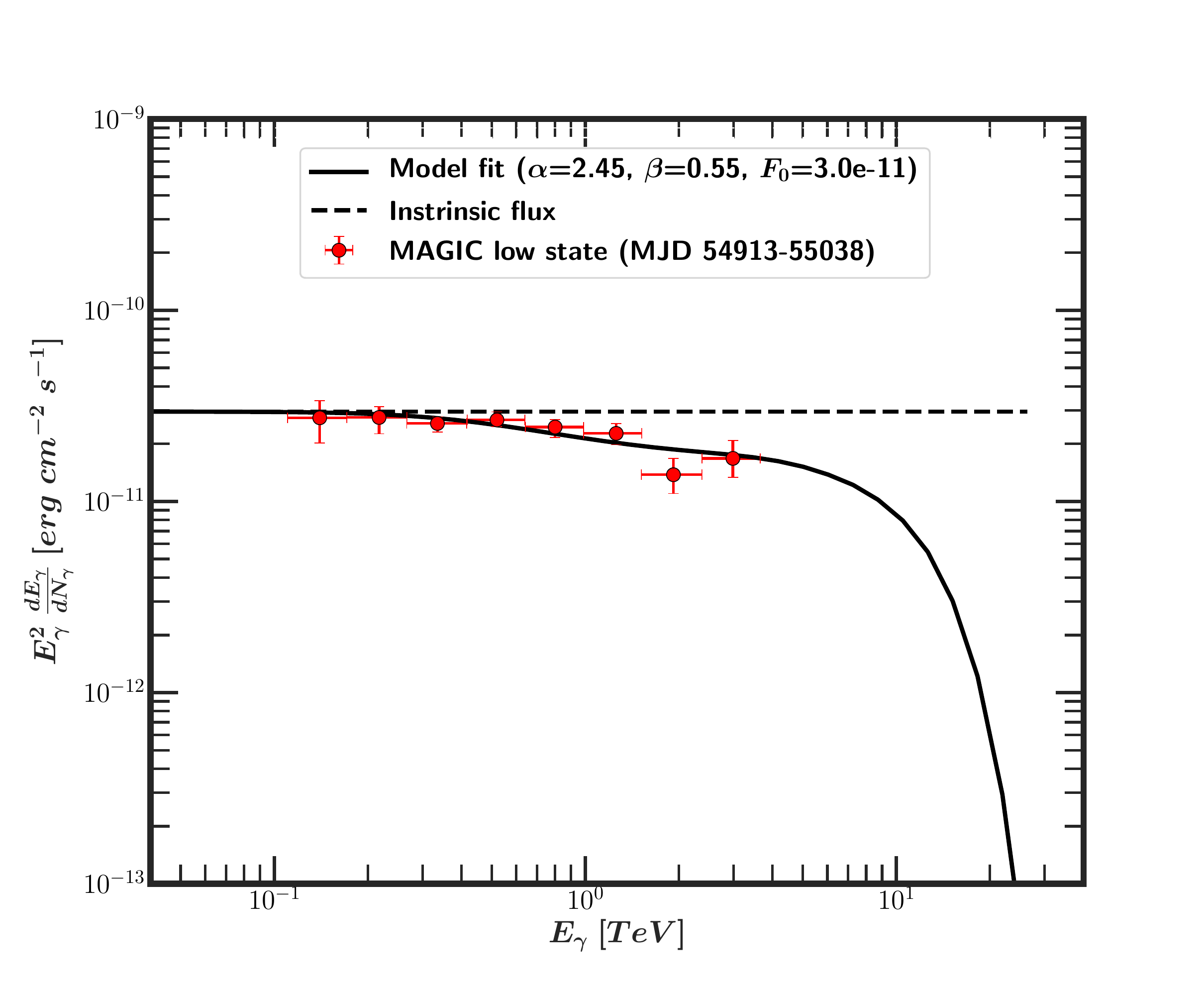}}
\par}
\caption{
MAGIC low state spectrum (MJD 54913 to MJD 55038) is fitted with the
photohadronic model. The corresponding intrinsic flux is also shown.
\label{fig:figure3}
}
\end{figure}

\section{Results}
As a part of multi-wavelength campaign, an extensive study of the HBL
Mrk 501 was conducted over 4.5 months prior in 2009, with the
participation of Whipple 10m, VERITAS, MAGIC, and many other
instruments\cite{Ahnen:2016hsf}. Although relatively large variability was measured in the
VHE $\gamma$-ray and X-ray bands, overall no significant correlation was
found between these energy bands. During the 4.5 months campaign period, the
VHE spectra were measured in three different states: low, high
and very high emission states. 

Here we shall analyze these three different states of Markarian 501
which are given in Table \ref{table1} by
using the photohadronic scenario discussed above. 
Here the EBL model of
Franceschini et al.\cite{Franceschini:2008tp} is used for the EBL correction.
It is observed that the EBL model of Dom\'inguez
also gives similar results\cite{Dominguez:2013lfa}.   

\begin{figure}
{\centering
\resizebox*{0.8\textwidth}{0.5\textheight}
{\includegraphics{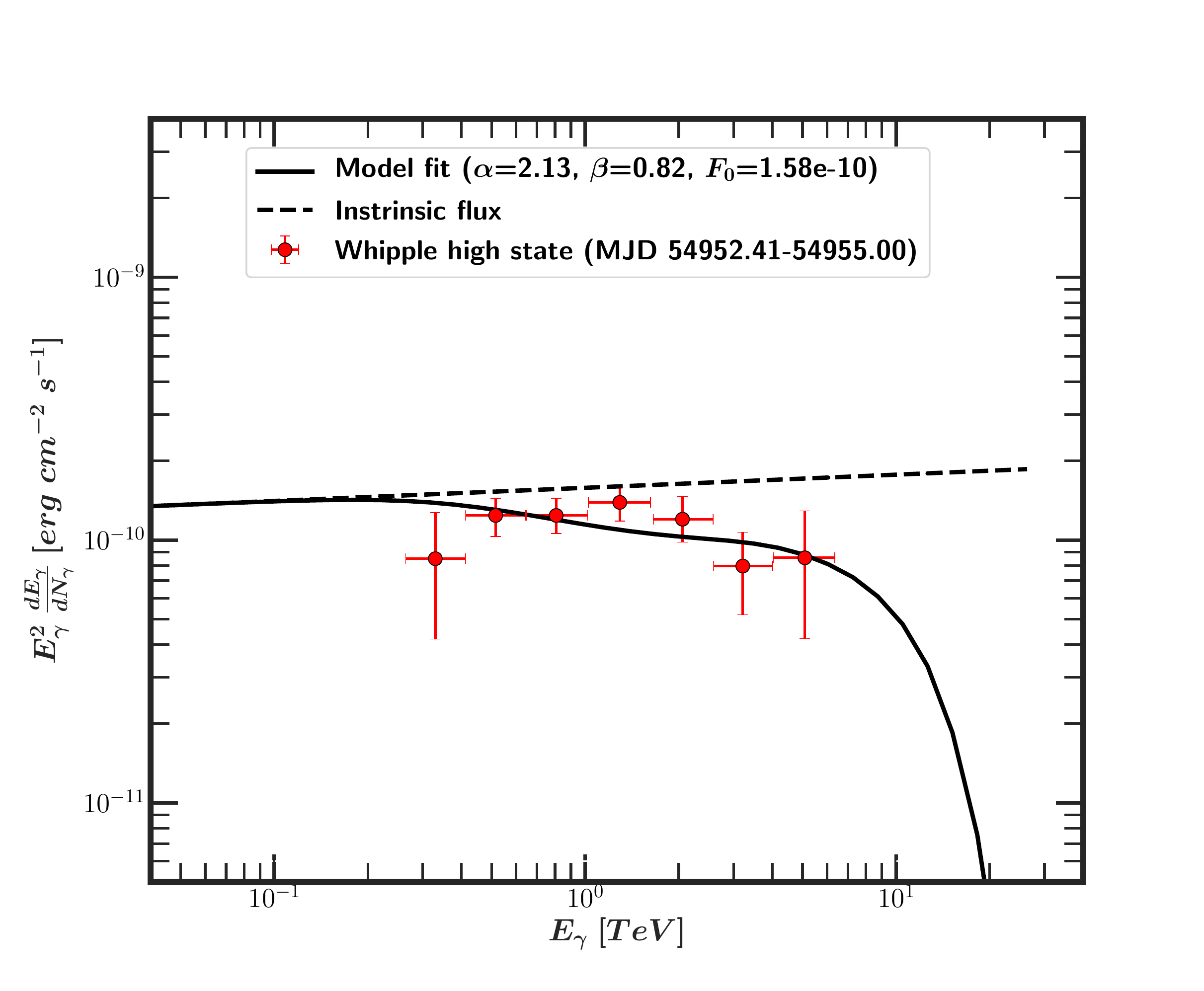}}
\par}
\caption{
The Whipple high state spectrum during MJD 54952.41 to MJD 54955.00 is fitted with the
photohadronic model. The corresponding intrinsic flux is also shown.
\label{fig:figure4}
}
\end{figure}

\begin{figure}
{\centering
\resizebox*{0.8\textwidth}{0.5\textheight}
{\includegraphics{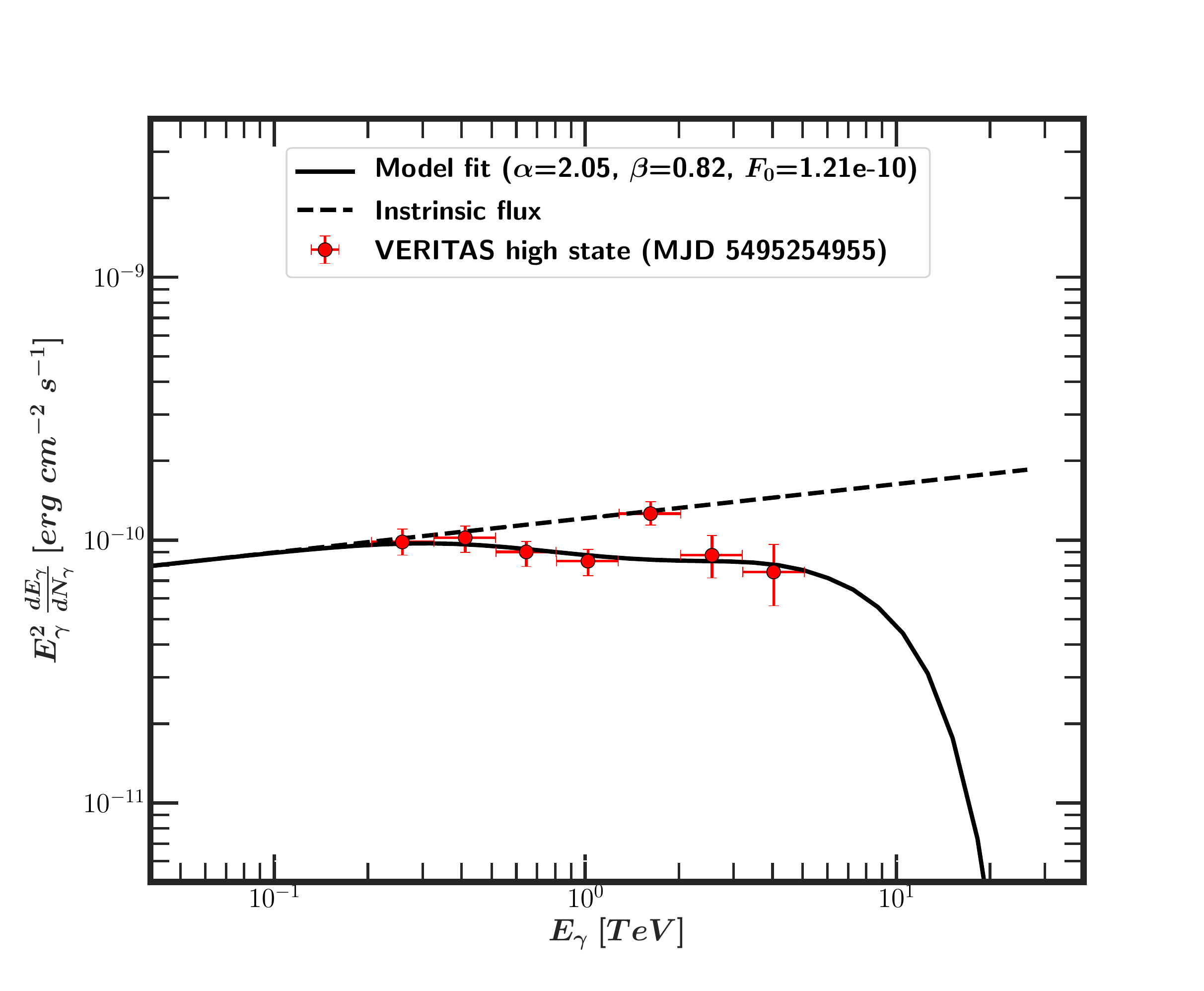}}
\par}
\caption{
The VERITAS high state spectrum observed during MJD 54952.41 to MJD
54955.00 is fitted using photohadronic model and the intrinsic
spectrum are shown in this figure.
\label{fig:figure5}
}
\end{figure}

\begin{figure}
{\centering
\resizebox*{0.8\textwidth}{0.5\textheight}
{\includegraphics{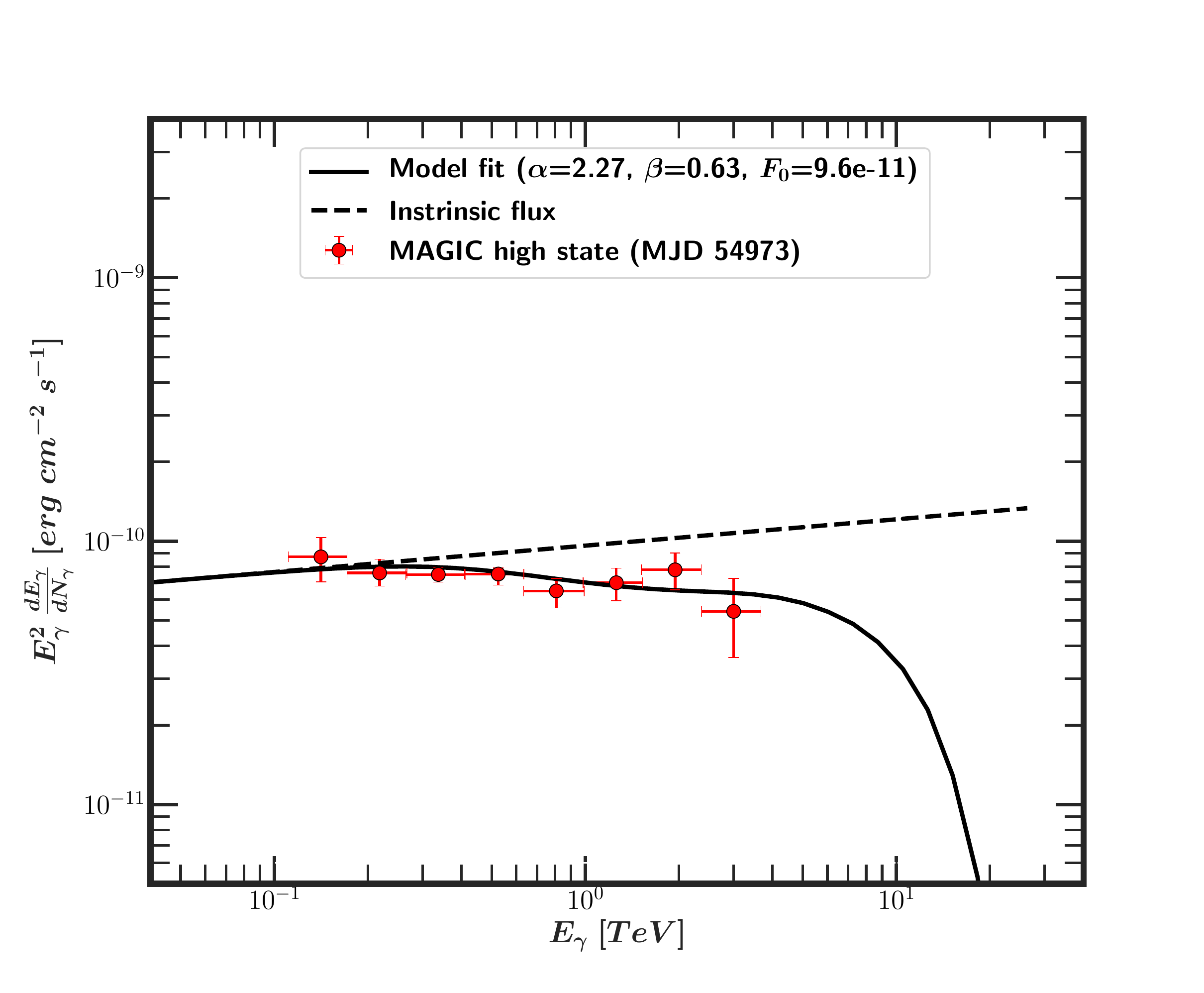}}
\par}
\caption{
MAGIC high state spectrum observed on MJD 54973 is fitted with the
photohadronic model. The corresponding intrinsic flux is also shown.
\label{fig:figure6}
}
\end{figure}

\begin{figure}
{\centering
\resizebox*{0.8\textwidth}{0.5\textheight}
{\includegraphics{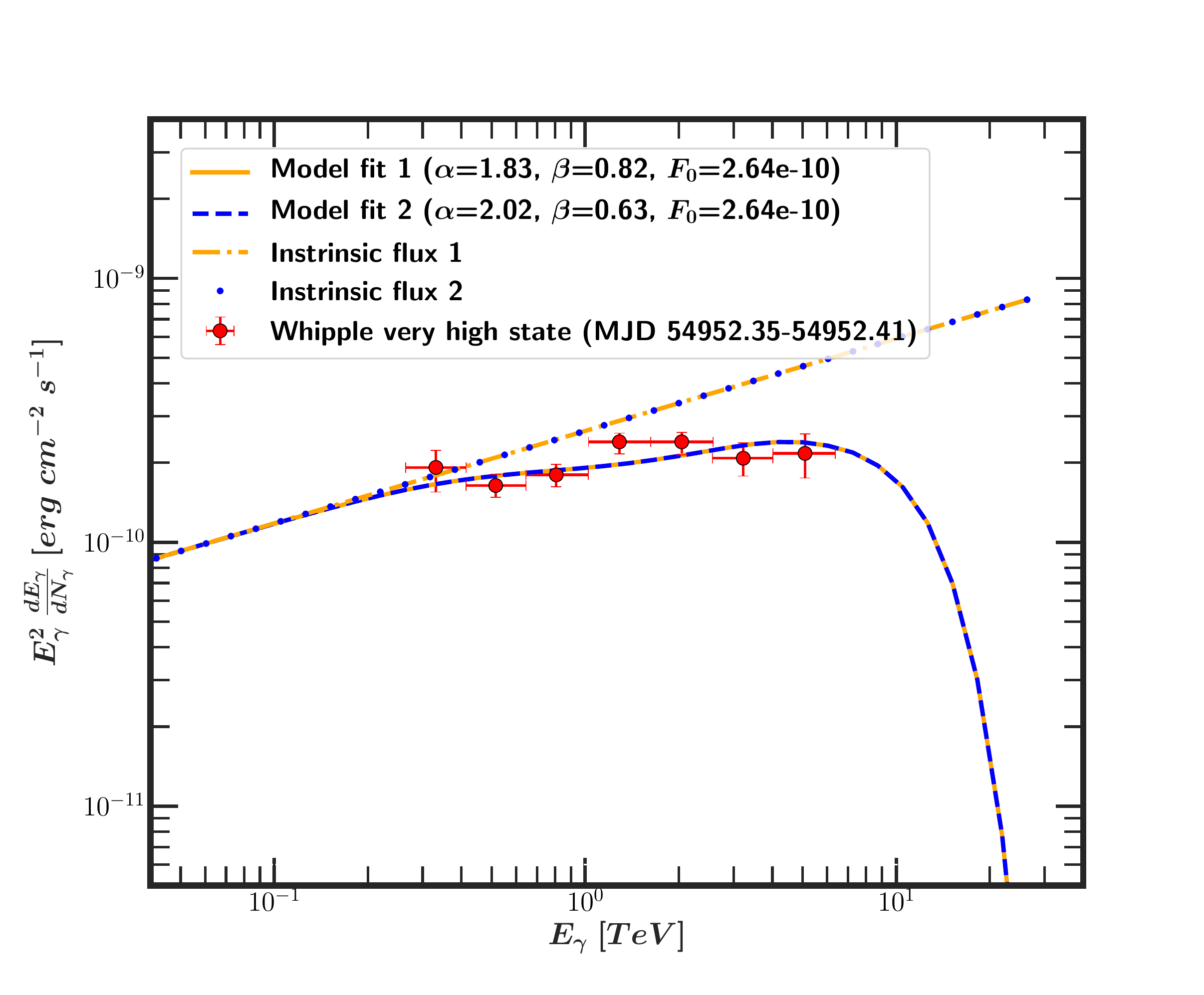}}
\par}
\caption{
The Whipple very high state flaring spectrum during MJD 54952.35 to MJD 54952.37 is fitted with the
photohadronic model. The corresponding intrinsic flux is also shown.
\label{fig:figure7}
}
\end{figure}

\begin{figure}
{\centering
\resizebox*{0.8\textwidth}{0.5\textheight}
{\includegraphics{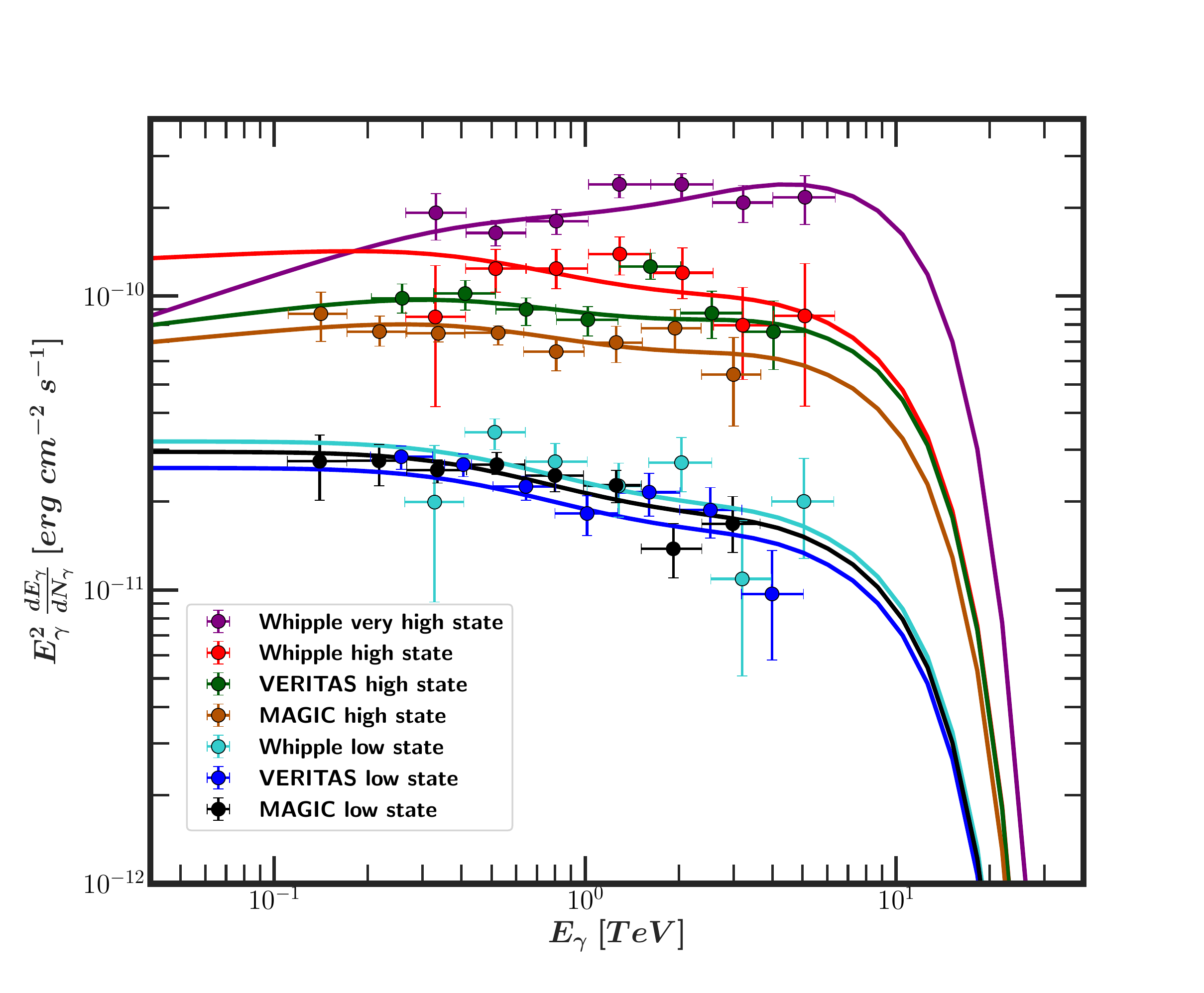}}
\par}
\caption{
Photohadronic fit to the low, high and very high emission states are
shown together along with the observed data for comparison.
\label{fig:figure8}
}
\end{figure}

\subsection{Low state}

The low states of Whipple, VERITAS and MAGIC are discussed here. During
mid-March to early-August 2009, the average MWL SED is taken when Mrk
501 was mostly in a low state\cite{Baring:2016pjj}. However, for
the calculation of these low states, the flaring events during the
observation period are excluded from the data. In the photohadronic
scenario, we need to know the tail region of the SSC spectrum which
will be used to fit the VHE spectrum and to calculate the spectral index $\alpha$. The SSC tail
region can be fitted with a power-law with
$\Phi_0=1.6\times\times 10^{-12}\, erg\, cm^{-2}\, s^{-1}$ and
$\beta=0.55$. These values of $\Phi_0$ and $\beta$ are used 
to fit the low state spectra of  Whipple, VERITAS and MAGIC and
$\Gamma=15$ is used here.

\subsubsection{Whipple} 
The Whipple low state spectrum between the period MJD 54936 to 54951
is observed in the energy range $0.33\, TeV\le E_\gamma \le 5.1 \,
TeV$ corresponding to the seed photon energy in the range 
$1.3\, MeV \ge \epsilon_\gamma \ge 20.5\, MeV$. The low state spectrum
is fitted very well with the following parameters $\alpha=2.45$ and
$F_0=3.2\times 10^{-11}\, erg\, cm^{-2}\, s^{-1}$. The fitted observed
spectrum (continuous curve) and the intrinsic spectrum (dashed curve)
are shown in Fig. \ref{fig:figure1}. 

\subsubsection{VERITAS}
The low state was
observed between MJD 54907 and MJD 55004 by VERITAS telescopes in the energy range 
$0.26\, TeV\le E_\gamma \le 4.0\, TeV$ which corresponds to the
SSC photon energy  $1.7\, MeV \ge \epsilon_\gamma \ge 26.3\, MeV$. A very
good fit to the data is obtained for the same $\alpha$ (2.45) as Whipple but 
$F_0=2.6\times 10^{-11}\, erg\, cm^{-2}\, s^{-1}$. The observed
spectrum, fitted curve and the intrinsic spectrum are shown in Fig. \ref{fig:figure2}.

\subsubsection{MAGIC}
The time averaged low state spectrum taken during MJD 54913 to MJD
 55038 is in the energy range $0.14\, TeV \le E_\gamma \le  3.0\,
 TeV$. To produce $\Delta$-resonance this corresponds to the seed
 photon energy in the range $2.2\, MeV \le \epsilon_\gamma \le 47.7\,
 MeV$. Again a very good fit to the low state spectrum is obtained with
 $\alpha=2.45$ and the normalization constant is $F_0=3.0\times
 10^{-11}\, erg\, cm^{-2}\, s^{-1}$. We also 
show the observed spectrum, the fitted curve and the intrinsic
spectrum in Fig. \ref{fig:figure3}.

As shown above the low states follow the same power-law ($\alpha=2.45$) except
that their normalizations are different (i.e. different values of
$F_0$), and their intrinsic spectra are flat irrespective of the
emission period.

\subsection{High state}
The high states observed by these three telescopes are discussed
below.

\subsubsection{Whipple}
The Whipple high state between MJD 54952.41 and MJD 54955.00 was observed,
when the flux was about two times the base line flux. It was observed in the energy range
$0.33\, TeV\le E_\gamma \le 5.1\, TeV$. The average multiwavelength
SED during this period is fitted using one-zone leptonic model in
ref.\cite{Aliu:2016kzx} . Using $\Gamma=15$ we observed that for the
above range of $E_{\gamma}$, the SSC seed photon energy is in
the range $1.3\, MeV \le \epsilon_\gamma \le 20.4\, MeV$. In this
range of $\epsilon_\gamma$, the SSC spectrum is fitted with a power-law
where $\Phi_0=7.0\times 10^{-12}\, erg\, cm^{-2}\, s^{-1}$ and
  $\beta=0.82$. In this case to have a good fit the high state of Whipple, we
  obtain $\alpha=2.13$ and $F_0=1.58\times 10^{-10}\, erg\, cm^{-2}\,
    s^{-1}$. The intrinsic flux is almost flat ($F_{\gamma,in}\propto
    E^{0.05}_{\gamma}$). The data and the fitted curve are shown in Fig. \ref{fig:figure4}.

\subsubsection{VERITAS}
The VERITAS high state corresponds to the same time period as that of Whipple
i.e.  MJD 54952.41 to MJD 54955.00. However, the VERITAS average flux
is lower than the one observed by Whipple, which was observed in
high state in the energy range $0.26\, TeV \le E_\gamma \le  4.0\,
TeV$. Here we use the same SSC flux (i.e. $\beta=0.82$) to fit the
observed spectrum. A very good fit to the data is obtained for
$\alpha=2.05$ and $F_0=1.21\times 10^{-10}\, erg\, cm^{-2}\, s^{-1}$
which is shown in Fig. \ref{fig:figure5}.

\subsubsection{MAGIC}
The MAGIC telescopes measured a high flux ($\sim 3$ times the low flux) on
May 22nd (MJD 54973) in the energy range $0.14\, TeV \le E_\gamma \le  3.0\,
TeV$. However, the low energy SED was not observed strictly
simultaneously during that period. As compared to the very high energy flare of
May 1st, this flare had a variability time scale of days, so
non-simultaneous SED may not affect much in the determination of the
value of $\beta$. Using the grid-scan modelling of the data obtained
during the flaring episode around MJD 54973
the low energy SED is modelled in Fig. 10 of 
ref.\cite{Aliu:2016kzx}. For the present calculation
we  consider a bulk Lorentz factor $\Gamma=15$. The energy range
$0.14\, TeV \le E_\gamma \le  3.0\, TeV$ implies that the seed
photon energy should be in the range $47.67\, MeV \ge \epsilon_\gamma \ge 2.25\,
MeV$ and the corresponding proton energy is in the range $1.4\,
TeV \le E_p \le  30\, TeV$. For the above $\epsilon_\gamma$ range
the corresponding SSC flux is a power-law with $\beta=0.63$
and $\Phi_0=3.6\times 10^{-12}\, erg\, cm^{-2}\, s^{-1}$. We fit the
MAGIC high state spectrum with $\alpha=2.28$ and
$F_0=9.6\times 10^{-11}\, erg\, cm^{-2}\, s^{-1}$, which is shown in
Fig. \ref{fig:figure6}  along with its intrinsic flux. 

\subsubsection{Whipple very high state}

On May 1, 2009, Whipple 10m telescope registered a very high flaring event,
the flux increased by a factor $\sim 4$
in the first 0.5 h (MJD 54952.35-MJD 54952.37) of the
observation and afterwards it decreased but remained in an elevated state
during MJD 54953-55 while the flux was about twice the baseline
flux. The sudden rise of the flux in a short time period
implies that the emission region was very small. Particularly on MJD
54952 when there was sub-hour flux variability, strictly-simultaneous
observations in multiwavelength were lacking and any SED constructed
during this period seems to be inconclusive. 
It is important to mention here that, the very high state spectrum
given in Fig. 8 of ref.\cite{Aliu:2016kzx} and the recent one in
Fig. 4 of ref.\cite{Ahnen:2016hsf} are different even
though they were taken in the same period (MJD 54952.35-MJD
52952.41), and the former one has a lower flux.
We observe that these spectra have
different values of $\alpha$ for the same $\beta$.

Due to non-simultaneous observation of the SSC SED and rapid variability of
the very high state, we don't know how the tail region of the SSC flux behaves, hence the
value of $\beta$ is uncertain. The bulk Lorentz
factor $\Gamma$ during this rapid variability period is also unknown.
However, to explain the high state emissions, $\beta=0.82$ was used
for the period MJD 54952.41-54955 which was immediately after the very
high state. Another high state observed on MJD 54973 is explained with
$\beta=0.63$. So here we use both values of $\beta$ to fit the
observed very high state spectrum.
Very good fit to the spectrum is obtained for (1) $\alpha=2.02$, 
$\beta=0.63$, and $F_0=2.53\times 10^{-10}\, erg\,
cm^{-2}\, s^{–1}$, and (2) $\alpha=1.83$, $\beta=0.82$, and $F_0=2.53\times 10^{-10}\, erg\,
cm^{-2}\, s^{–1}$, which are exactly the same as shown in
Fig. \ref{fig:figure7}. As we know, the proton spectral index $\alpha$
should be $\ge 2$. However, for case (2) we have $\alpha < 2$
even though it fits very well to the data. Hence, we only consider the
first case for which the intrinsic flux  $F_{\gamma,in}\propto
E^{0.35}_{\gamma}$. It seems, the SSC SED during the very high state emission must have $\beta \le
0.63$.
For comparison we show the spectra of the low, the high, and the very high
emission states observed by different instruments and their respective
photohadronic fits in Fig. \ref{fig:figure8}.

In the flaring state, as has been alluded to before, in general the
flux of the two opposing jets can be as high as $F_{Edd}/2$. Mrk 501 has a central black hole of mass $M_{BH}=(0.9-3.5)\times
10^9\,M_\odot$ which corresponds to an Eddington luminosity
$L_{Edd}=(1.13-4.4)\times 10^{47}\, erg\, s^{-1}$ and
$F_{Edd}=(3.9-15.0)\times 10^{-8}\,  erg\, cm^{-2}\, s^{-1}$. The proton
flux $F_p$ corresponding to the highest $\gamma$-ray of $E_\gamma=5.09$
TeV must satisfy $F_p<F_{Edd}/2=2.0\times
10^{-8}\,erg\,cm^{-2}\,s^{-1}$. Using the relation between the proton
flux and the $\gamma$-ray flux we obtain $\tau_{p\gamma}>0.08$. For a
moderate efficiency of the $p\gamma$ process we take
$\tau_{p\gamma}=0.1$ for which we obtain $F_p=1.6\times 10^{-8}
\,erg\, cm^{-2}\, s^{-1}<F_{Edd}/2$, which shows that the highest
energy proton has sub-Eddington luminosity.

It is very important to note that during different VHE emission states 
from Mrk 501, the multiwavelength SEDs are different in
the SSC frequency range (different values of $\beta$) which is obvious from the leptonic model fit
to the SEDs. Putting in other words, this corresponds to different
seed photon densities in the SSC band in each epoch of flaring as
given in Eq. (\ref{photondensity}). But the proton
spectral index lies in a narrow range $2.02\lesssim\alpha\lesssim 2.45$ which
shows that the high energy proton acceleration mechanism is the same
for low, high and very high flaring states. We have observed that the
photohadronic model works well for $E_{\gamma} \gtrsim 100$ GeV and in
this energy range the SSC contribution is negligible.

\section{Conclusions}

The HBL Mrk 501 was observed during a multiwavelength campaign covering a period
of 4.5 months from March 15 to August 1, 2009 \cite{Ahnen:2016hsf}. In this period three
different types of VHE emissions were observed by Whipple 10m, VERITAS
and MAGIC telescopes. A very high state flaring event was observed only by
Whipple telescope for about 0.5 h when the flux had a dramatic
increase. All the three telescopes also observed high emission state
and low emission state of Mrk 501 during this campaign period. Using the
photohadronic scenario, where Fermi accelerated protons interacting
with the seed photons in tail region of the SSC SED in the inner jet
region produce the $\Delta$-resonance and its decays to neutral pion will subsequently
produce observed VHE photons. We have shown that all these three types
of VHE spectra can be fitted very well with the photohadronic scenario when
absorption by EBL is accounted for. Also we
observed that the intrinsic spectra of these three different states are
different from each other. The low state spectrum is almost flat 
which shows that $F_{\gamma,in} $ is constant and the one from the
high state is $\propto E^{0.1}_{\gamma}$. The $F_{\gamma,in} $ from
the very high state is found to be proportional to
$E^{0.35}_{\gamma}$. So going from low state to very high state, the
intrinsic  flux slowly increases from a constant to a power-law,  
and simultaneously the observed flux follows the trend from downward
going to flat to upward going Fig. (\ref{fig:figure8}).
We also observed that the proton spectral index for all these
cases lies in a small window of $2.02\lesssim \alpha \lesssim 2.45$. It seems that
for low, high, and very high state emissions from Mrk 501, the 
acceleration mechanism for the high energy proton is the same,
but the only difference is coming from the seed photon density
(different values of $\beta$). So it is important that we should have
simultaneous multiwavelength observations of the flaring event to
constrain the photohadronic model.

\acknowledgements

The  work of S.S. is partially supported by DGAPA-UNAM (M\'exico) Project
No. IN103019 and PASPA-DGAPA, UNAM. S.S. is also thankful to Japan Society for the Promotion
of Science (JSPS) for the invitational fellow program.

\end{document}